\documentclass[unnumsec,webpdf,modern,large]{mam-authoring-template}%

\graphicspath{{Fig/}}

\theoremstyle{thmstyleone}%
\theoremstyle{thmstyletwo}%
\theoremstyle{thmstylethree}%

\newcommand{\vect}[1]{\ensuremath{\vec{#1}}}
\usepackage{gensymb}
\usepackage{hyperref}
\newcommand{\suppl}[1]{\href{https://doi.org/10.48550/arXiv.2411.15323}{#1}} 
\usepackage{lmodern}

\begin{document}

\journaltitle{Microscopy and Microanalysis}
\DOI{\href{https://doi.org/10.48550/arXiv.2411.15323}{10.48550/arXiv.2411.15323}}
\copyrightyear{2025}
\pubyear{2025}
\appnotes{Preprint}

\firstpage{1}


\title[EH-VFT magnetisation reconstruction]{Model-Based Iterative Reconstruction of Three-Dimensional Magnetisation in a Nanowire Structure Using Electron Holographic Vector Field Tomography}

\author[1,$\ast$]{Aurys \v{S}ilinga\ORCID{0000-0001-6124-1754}}
\author[2]{Andr\'as Kov\'acs\ORCID{0000-0001-8485-991X}}
\author[1]{Stephen McVitie\ORCID{0000-0003-4511-6413}}
\author[2]{Rafal E. Dunin-Borkowski\ORCID{0000-0001-8082-0647}}
\author[1]{Kayla Fallon}
\author[1,$\ast$]{Trevor P. Almeida\ORCID{0000-0003-4683-6279}}

\authormark{Aurys \v{S}ilinga et al.}

\address[1]{\orgdiv{School of Physics and Astronomy}, \orgname{University of Glasgow}, \orgaddress{\street{Kelvin Building}, \postcode{Glasgow G12 8QQ}, \state{Scotland}, \country{United Kingdom}}}
\address[2]{\orgdiv{Ernst Ruska-Centre for Microscopy and Spectroscopy with Electrons}, \orgname{Forschungszentrum J\"ulich GmbH}, \orgaddress{\street{Wilhelm-Johnen-Stra{\ss}e}, \postcode{52428 J\"ulich}, \state{North Rhine-Westphalia}, \country{Germany}}}

\corresp[$\ast$]{Corresponding authors: \href{email:email-id.com}{a.silinga.1@research.gla.ac.uk}, \href{email:email-id.com}{trevor.almeida@glasgow.ac.uk}}


\abstract{Experimental techniques for the characterisation of three-dimensional (3D) magnetic spin structures are required to advance the performance of nanoscale magnetic technologies. However, as component dimensions approach the nanometre range, it becomes ever more challenging to analyse 3D magnetic configurations quantitatively with the required spatial resolution and sensitivity. Here, we use off-axis electron holography and model-based iterative reconstruction to reconstruct the 3D magnetisation distribution in an exemplary nanostructure comprising an L-shaped ferromagnetic cobalt nanowire fabricated using focused electron beam induced deposition. Our approach involves using off-axis electron holography to record tomographic tilt series of electron holograms, which are analysed to reconstruct electron optical magnetic phase shifts about two axes with tilts of up to $\pm60\degree$.  A 3D magnetisation vector field that provides the best fit to the tomographic phase measurements is then reconstructed, revealing multiple magnetic domains in the nanowire. The reconstructed magnetisation is shown to be accurate for magnetic domains that are larger than approximately 50~nm. Higher spatial resolution and improved signal-to-noise can be achieved in the future by using more specialised electron microscopes, improved reconstruction algorithms, and automation of data acquisition and analysis. \newline  \newline  \includegraphics[width=14.6cm]{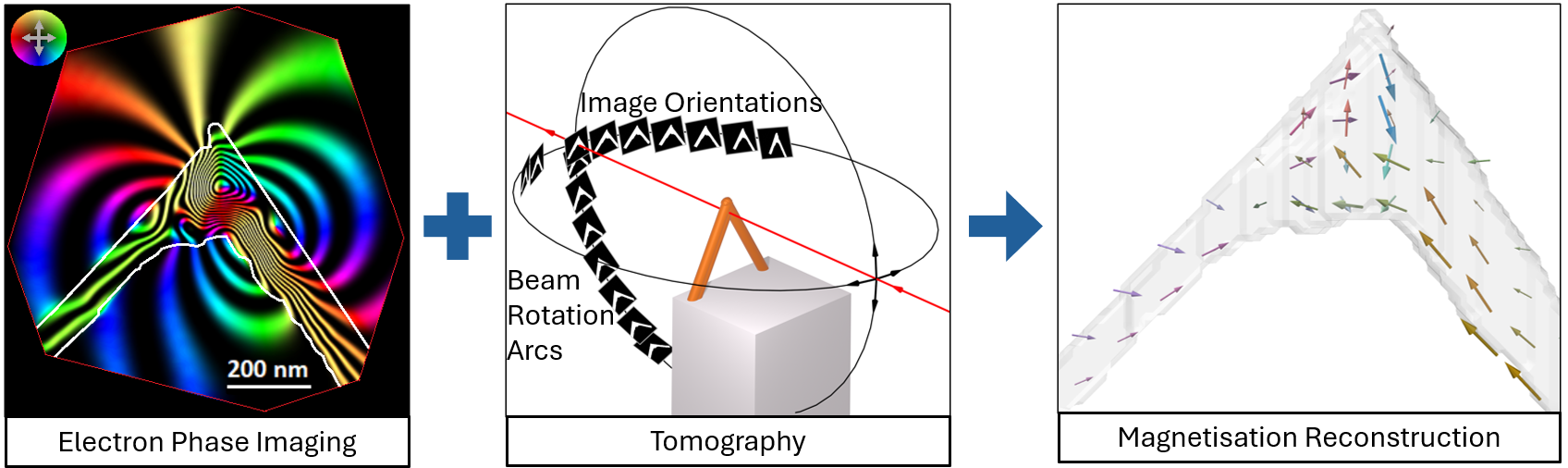}}

 

\maketitle

\section{Introduction}\label{section1}

\subsection{3D magnetic characterisation}\label{subsec11}

As a result of advances in material characterisation, fabrication and the discovery of increasingly complex spin textures such as skyrmions and hopfions \citep{zheng_hopfion_2023}, there has been a drive to explore the properties of three-dimensional (3D) magnetic configurations. Spintronic circuits that contain integrated magnetic nanowires have been proposed for next-generation data storage applications \citep{parkin_magnetic_2008} and as physical components for improving machine-learning efficiency \citep{ellis_machine_2023}. Spintronic computing is being developed due to its energy efficiency compared to capacitor-based electronics \citep{nikonov_power_2006}. However, most prototypes have been fabricated using two-dimensional (2D) lithographic techniques and the 3D stacking that would benefit commercial implementation has not been optimised. For example, magnetic racetrack memory (RM) prototypes have demonstrated storage and manipulation of data carried by magnetic domains in 2D patterned nanowires. Their extension into 3D nanowire architectures would increase the storage density per unit area \citep{gu_three-dimensional_2022}. As prototypes of 3D magnetic architectures are developed, experimental measurements of magnetic structure and domain motion are required to guide further development of RM-type technologies. The observables that are used to characterise ferromagnetic samples are related by the equation
\begin{equation}
\vect{B} = \mu_{0} (\vect{H}(\vect{M})+\vect{M})~,
\label{eq1}
\end{equation}	
where $\vect{M}$ is the magnetisation vector field describing magnetic moment per unit volume, \vect{B} is the magnetic induction and $\mu_0$ is the vacuum permeability. If no external magnetic fields, conduction currents or displacement currents are present, then  $\vect{H}$ is the demagnetising field, which is defined by the distribution of $\vect{M}$. \textcolor{black}{   It should be noted that, in general, \vect{B} permeates all of space, while \vect{M} is local to the sample and has a more compact description than \vect{B}.} X-ray imaging currently allows 3D reconstruction of $\vect{M}$ with sub-50-nm resolution using magnetic laminography \citep{donnelly_time-resolved_2020} and sub-70-nm resolution using Fourier transform holography \citep{donnelly_three-dimensional_2017,di_pietro_martinez_three-dimensional_2023}. Theoretical predictions estimate a resolution for X-ray laminography down to 20~nm \citep{donnelly_complex_2022}. In comparison, phase contrast techniques in the transmission electron microscope (TEM) cannot be used  to reconstruct $\vect{M}$ directly, but have been used to measure \vect{B} with a 3D resolution of 10~nm using electron holographic vector field tomography (EH-VFT) \citep{phatak_vector_2008,phatak_three-dimensional_2010,wolf_holographic_2019}. Micromagnetic simulations have also been used to interpret TEM measurements of \vect{B} \citep{donahue_exchange_1997,lyu_three-dimensional_2024}. Here, we use the alternative approach of applying model-based iterative reconstruction (MBIR) to reconstruct a magnetisation distribution \vect{M} from experimental \textcolor{black}{TEM} measurements that are sensitive to $\vect{B}$ rather than \vect{M}. We discuss how magnetic and electrostatic signals obtained from electron phase measurements can be used to improve the reconstruction, as well as practical limitations to resolution and accuracy. \textcolor{black}{Whereas most previous TEM studies have used EH-VFT to reconstruct $\vect{B}$ in 3D, we demonstrate the reconstruction of $\vect{M}$  in 3D by applying MBIR to an EH-VFT dataset.}

\subsection{Model-based iterative reconstruction of \vect{M}}\label{subsec12}

The theoretical development of MBIR for the tomographic reconstruction of \vect{M} from EH-VFT measurements has been described in detail elsewhere  \citep{caron_thesis_2018}. The most common phase contrast techniques for magnetic imaging in the TEM \citep{lubk_differential_2015,kohn_experimental_2016,krajnak_pixelated_2016,you_lorentz_2023,cui_antiferromagnetic_2024}, which are often collectively termed Lorentz microscopy, all rely fundamentally on an interpretation of the Aharonov-Bohm effect \citep{aharonov_significance_1959}, which describes how the phase of an electron wave is affected by electromagnetic potentials. The interaction of an electron wave with a magnetic induction field \vect{B}  results in an electron phase shift $\varphi_{mag}$, where 
\begin{equation}
\int{\vect{B} \times \text{d} \vect{z}= \frac{\hbar}{e}\vect{\nabla} \varphi_{mag}~,}
\label{eq2}
\end{equation}
$e$ is the electron charge, $\hbar$ is the reduced Planck constant, $\text{d} \vect{z}$ is a path element in the incident electron beam direction and the phase gradient $\vect{\nabla} \varphi_{mag}$ is defined in a 2D plane perpendicular to the incident electron beam direction. In principle, MBIR can be applied to results obtained using all Lorentz microscopy techniques, as they are all sensitive to $\varphi_{mag}$. Here, the recovery of $\varphi_{mag}$ from off-axis electron holograms is described in the Methods section.
Typically, \vect{H} fields are present and therefore \vect{B} and \vect{M} are not equivalent. For imaging conditions where \vect{H} results only from the \vect{M} configuration, magnetic phase effects are considered to arise from the $\text{curl}(\vect{M})$ component parallel to the electron beam direction \citep{mcvitie_imaging_2003}. This $\text{curl}(\vect{M})$ component, which can be regarded as an Amperian current, is the origin of the \vect{B} component that contributes to the electron phase and is associated only with the sample. The \vect{M} component of magnetic vortex domain walls (DWs) has been measured under such conditions \citep{junginger_quantitative_2008}, but it required tilting of the sample to evaluate all of the components of $\text{curl}(\vect{M})$. The reconstruction error depends on the type of sample, the angular imaging range and the correction of TEM instrument misalignments. If a nanostructure can be tilted incrementally to high angles and $\varphi_{mag}$ is detected from all parts of the sample, then all the information that is needed for reconstruction can be obtained from the phase measurements. Simulation-based error estimates for this experiment are discussed in the Methods section and the Supplementary Material \suppl{[S1]}.

\begin{figure*}[t]
        \centering\includegraphics[width=16cm]{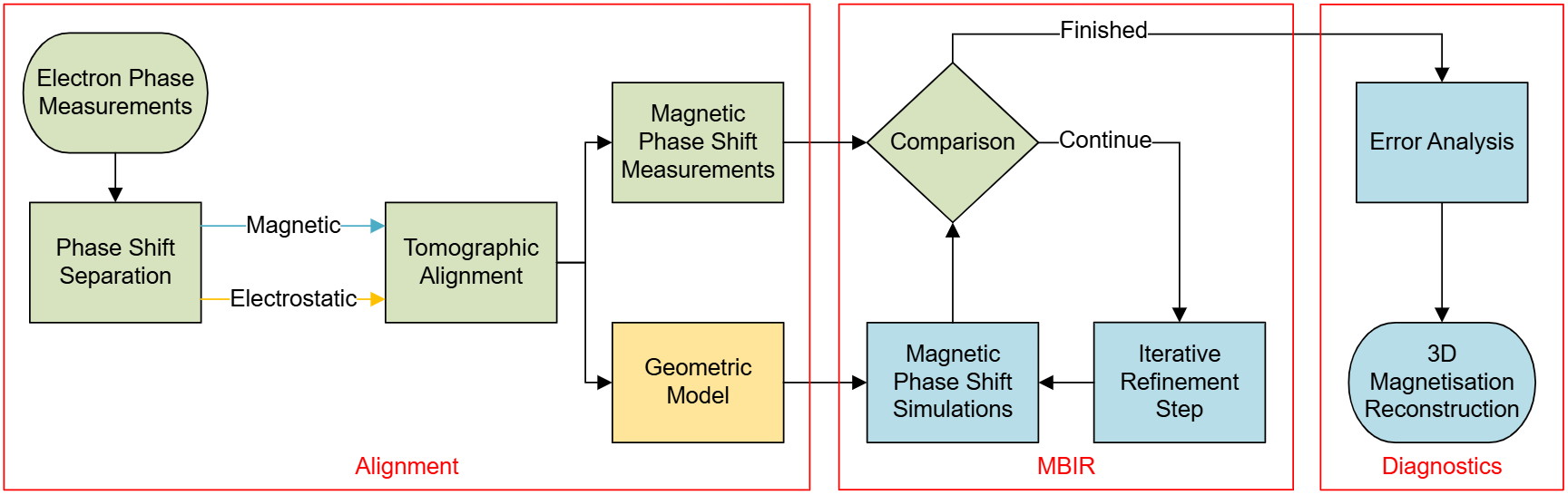}
        \caption{Workflow for 3D magnetisation reconstruction. The red boxes indicate the primary processing steps: alignment, reconstruction using MBIR, and diagnostics. The steps are shaded to indicate the data type that is being processed: two-dimensional phase images (green), 3D magnetic vector fields (blue) and a 3D electrostatic scalar field (yellow). Datasets are processed to generate both a geometrical mask that defines where the magnetic material is located and a series of magnetic phase shift measurements corresponding to \vect{B} field projections. At each iteration, a distribution of \vect{M} is generated, and then its magnetic phase shift is simulated and compared to the measurements. The iterations are repeated until an optimal \vect{M} is found. Optimal estimation diagnostics are performed to assess random and systematic errors in the reconstruction. The reconstruction is improved by accurate alignment of the experimental data.}
        \label{fig_workflow}
\end{figure*}

The workflow for the reconstruction of \vect{M} (Fig.~\ref{fig_workflow}) includes: 1)~Processing and alignment of experimental data to construct a 3D geometrical model of the sample; 2)~MBIR reconstruction of the \vect{M} distribution that best matches the experimental data; 3)~Optimal estimation diagnostics for evaluating reconstruction errors. The theory of applying MBIR for \vect{M} reconstruction has been developed and tested extensively using simulated datasets \citep{caron_chapter_6}. The present work is an experimental development, which addresses several sources of error, such as those caused by sample alignment \citep{diehle_cartridge-based_2021}, sample drift, and surface damage. It also builds on other previous work on the 3D reconstruction of \vect{M} from experimental electron microscopy data \citep{mohan_model-based_2018}. 

\subsection{Null spaces in electron phase measurement}\label{subsec13}

In Lorentz microscopy, some configurations of \vect{M} \citep{mansuripur_computation_1991} contribute no signal to experimental measurements made in projection and can be described as null spaces, where \vect{M} may be reconstructed falsely as being close to zero \citep{caron_chapter_4_5}. In such projections, the vector components of \vect{M} can sum to zero or have non-zero divergence of \vect{M}. In some cases, tiltng the sample can reveal the hidden components. However, divergent components can remain undetectable even when tilting of the sample is used. As a result, some configurations of \vect{M}, such as N\'eel-type magnetic domain walls, cannot be reconstructed fully \textcolor{black}{because they contain some states of \vect{M} that do not create a detectable TEM signal in any projection. It has been observed experimentally \citep{mcvitie_transmission_2018} and proven mathematically  \citep{caron_chapter_4_5} that N\'eel-type magnetic domain walls in both out-of-plane and in-plane magnetised films include eigenstates of the \vect{M} distribution that do not result in electron phase shift. Such states cannot be reconstructed without making use of additional constraints, and examples of the application of MBIR to such states are discussed in the Supplementary Material \suppl{[S2]}.} It should be noted that null spaces are not present in the experimental reconstruction described in this paper.

\section{Materials and methods}\label{section2}
\subsection{Sample fabrication and characterisation}\label{sebsec21}

\begin{figure*}[t]
        \centering\includegraphics[width=12cm]{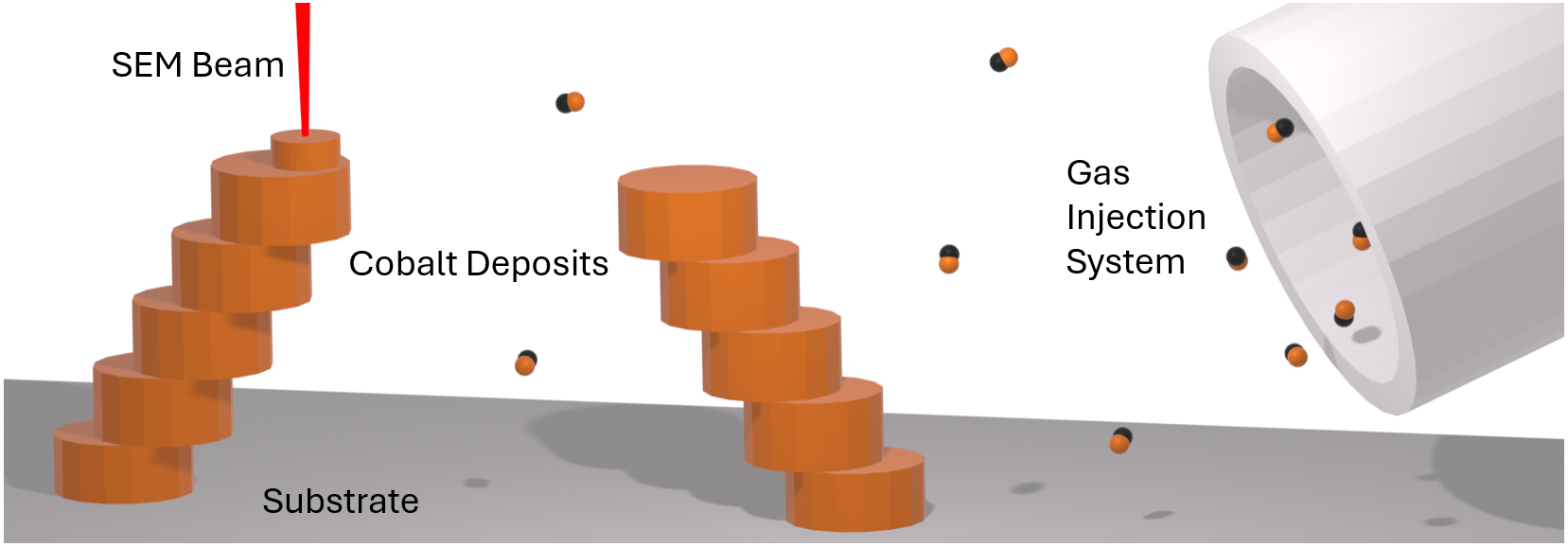}
        \caption{FEBID in a scanning electron microscope (SEM). Cobalt is deposited locally where the electron beam interacts with solid material. The electron beam is translated to control the location of the deposited material.}
        \label{fig_febid}
\end{figure*}

\begin{figure*}[t]
        \centering\includegraphics[width=14cm]{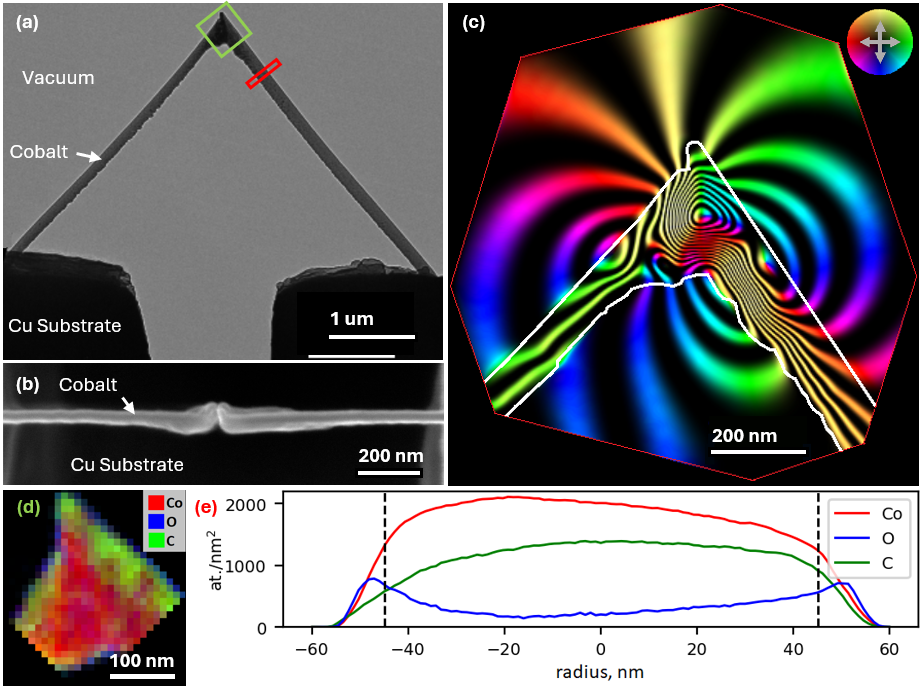}
        \caption{Dimensions, morphology and chemical analysis of an L-shaped cobalt nanostructure fabricated using FEBID.(a)~Bright-field TEM side-view image. The marked regions correspond to the elemental map (green) and cross-sectional line profile (red) shown in (d) and (e), respectively. (b)~SEM top-down image of the tip of the sample. (c)~Magnetic phase contour image recorded using off-axis electron holography, displayed as $\text{cos}(6 \varphi_{mag})$. The colours show the direction of the projected in-plane \vect{B} field. The sample outline is marked. (d)~EELS elemental map of the tip, showing the distributions of Co, O and C. (e)~Cross-sectional EELS elemental line profile, showing that the central 90~nm of the nanowire has a cobalt content of between 45\% and 60\%.}
        \label{fig_sample}
\end{figure*}

An L-shaped intersection between ferromagnetic cobalt nanowires was studied, as it is similar to an element of a 3D racetrack memory configuration, in which structural features can act as magnetic domain wall pinning sites. In order to permit TEM imaging of magnetic states at such pinning sites, free-standing cobalt nanowires were fabricated on a copper substrate using focused electron beam deposition (FEBID) in a scanning electron microscope (SEM) \citep{skoric_layer-by-layer_2020, magen_focused-electron-beam_2021}. As shown in Fig.~\ref{fig_febid}, a precursor gas is injected into the SEM chamber and cobalt is deposited at locations where the electron beam interacts with the sample. The geometry of the deposited structure is controlled by translating the electron beam in a controlled manner using the SEM scan coils. The structure shown in Fig.~\ref{fig_sample}a, which consists of two intersecting nanowires, was defined using computer-assisted design (CAD) \citep{skoric_layer-by-layer_2020} and deposited by electron irradiation of a Co$_2$(CO)$_8$ precursor in an FEI HELIOS Plasma focused ion beam SEM. Deposition was performed using a 30~kV electron beam with a 690~pA beam current and a 6$\times10^{-5}$~Pa chamber deposition pressure. The dimensions and morphology of the sample were determined by imaging the intersection from multiple tilt angles in the SEM, such as the top-down image show in Fig.~\ref{fig_sample}b. Scanning TEM electron energy-loss spectroscopy (EELS) was used for chemical analysis of the green and red boxed regions marked in Fig.~\ref{fig_sample}a to obtain an elemental map (Fig.~\ref{fig_sample}d) and cross-sectional line profiles (Fig.~\ref{fig_sample}e). The EELS results show that the nanowire core has a cobalt purity in the range of 45\% to 60\%, while the surface is covered in a shell of carbon and oxygen. The radial dependence of the cobalt content in the cross-sectional elemental map shown in Fig.~\ref{fig_sample}e suggests that the deposition was performed in the beam-limited FEBID r\'egime \citep{serrano-ramon_ultrasmall_2011}, in which the material that receives the highest radiation dose during deposition has the highest purity. Off-axis electron holography was performed to image the local magnetic configuration of the sample. Prior to imaging in the TEM, the sample was saturated magnetically by applying a 1~T magnetic field in the plane of the nanowire perpendicular to the substrate and relaxed into the magnetic state shown in Fig.~\ref{fig_sample}c. The magnetic contribution to the phase image records the gradient of the projected in-plane components of \vect{B}, according to Eq.~\ref{eq2}. It is affected by local variations in cobalt purity, projected sample thickness and the orientation of $\text{curl}(\vect{M})$. As the present sample contains a head-to-head magnetic domain wall with a vortex configuration, tomographic reconstruction is needed for full magnetic characterisation.

\subsection{Electron holographic vector field tomography}\label{subsec22}

\begin{figure*}[t]
        \centering\includegraphics[width=15cm]{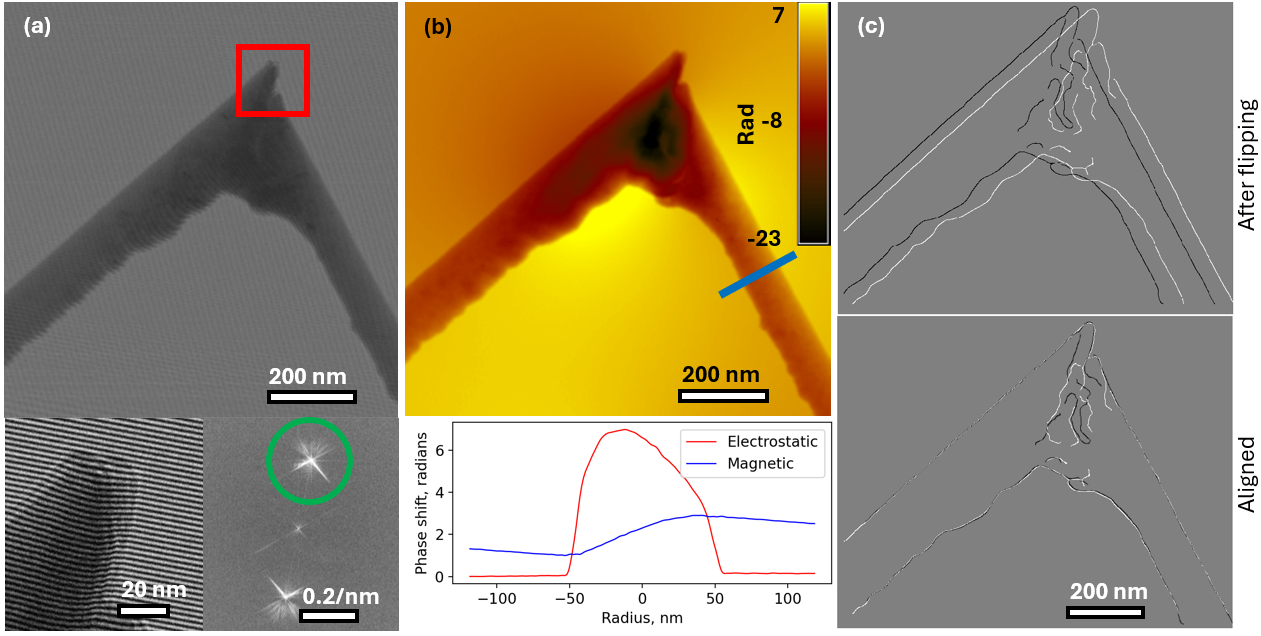}
        \caption{Off-axis electron holograms and alignment of phase images. (a)~Off-axis electron hologram formed by the interference of two electron waves, one of which has passed through the region of interest on the sample. The insets show a magnified version of the region inside the red box (lower left) and an FFT of the electron hologram (lower right). The green circle in the FFT marks the sideband that was used to used to calculate~(b). (b)~Total phase image obtained from an inverse FFT of the sideband. The inset shows a line profile of the electrostatic and magnetic contributions to the total phase shift along the blue line. (c)~Steps in the alignment of phase images recorded before and after flipping the sample by 180$\degree$ to separate the electrostatic and magnetic contributions to the phase. Image distortions resulting in misalignment (top) are corrected by applying an affine transformation (bottom).}
        \label{fig_holo}
\end{figure*}

In this work, off-axis electron holography \citep{dunin-borkowski_electron_2019} is chosen as the preferred magnetic characterisation technique because it measures the electron phase shift directly and has a high phase sensitivity of 0.016~radian root-mean-square noise for a 10~s exposure time per phase image. It also causes minimal sample damage when using a low electron flux of 200~e/nm$^2$/s. Tilt series of off-axis electron holograms for EH-VFT were recorded using an FEI Titan G2 60-300 TEM at 300~kV. A voltage of 100~V was applied to the biprism, resulting in electron holograms (Fig.~\ref{fig_holo}a) with an interference fringe spacing of $\sim 3.1$~nm, as shown in the left inset to Fig.~\ref{fig_holo}a. The phase shift (Fig.~\ref{fig_holo}b) was determined from the inverse fast Fourier transform (FFT) of one of the sidebands of each hologram, as shown in the right inset to Fig.~\ref{fig_holo}a \citep{mitchell_scripting-customised_2005,boureau_off-axis_2018,dunin-borkowski_electron_2019}. At each sample position, five holograms with individual exposure times of 2~s were recorded and the corresponding phase images were averaged to reduce the effects of statistical noise. A large TEM pole-piece gap ($\sim 11$ mm) and a dedicated tomography holder \citep{diehle_cartridge-based_2021} allowed the sample to be tilted by 360$\degree$ about the holder axis without removing it from the microscope. 

Tilt series of off-axis electron holograms were recorded at 10$\degree$ tilt increments over two \textcolor{black}{orthogonal} arcs. The effective maximum tilt angles were limited to between -60$\degree$ and 30$\degree$ and to between -60$\degree$ and 0$\degree$, as the sample was shadowed by the holder and the substrate. \textcolor{black}{For each arc, holograms were recorded at the same tilt angles before and after turning the sample over, in order to enable phase shift separation, which is described below.} The second arc was recorded to reduce missing wedge artefacts and to prevent projection-based null spaces \citep{caron_chapter_4_5}. Simulations of samples comprising uniformly-magnetised nanowires and vortex configurations have shown that the use of smaller tilt increments has a negligible impact on reconstruction quality \citep{caron_chapter_6}. Furthermore, simulations presented in the Supplementary Material \suppl{[S1]} show that if a 10\% difference between the volumes of the sample and its geometrical model is assumed then the difference between the reconstruction of \vect{M} and the ground truth is on average less than 10\% per voxel.
In order to separate the electrostatic and magnetic contributions to the phase images, off-axis electron holograms were recorded before and after turning the sample by 180$\degree$. This approach changes the sign of the magnetic contribution to the total phase shift, based on the expression \citep{aharonov_significance_1959,weyland_electron_2016},
\begin{equation}
    \begin{gathered}
    \varphi(x,y)=\varphi_{el}(V) + \varphi_{mag}(\vect{A}) \\
 = \frac{e}{\hbar \nu}\int V(x,y,z) \text{d}z + \frac{e}{\hbar}\int \vect{A}(x,y,z) \cdot \text{d}\vect{z}~,
    \label{eq3}
    \end{gathered}
\end{equation}
where $\varphi_{el}(V)$ and $\varphi_{mag}(\vect{A})$ are the electrostatic and magnetic contributions to the phase shift, respectively,  $\nu$ is the relativistic electron speed and $\text{d}z$ and $\text{d}\vect{z}$ are scalar and vector elements of the electron path, respectively. For a pair of such reversed holograms, half of the sum and half of the difference of the measured phase images represent  $\varphi_{el}$ and $\varphi_{mag}$, respectively. Unfortunately, physically turning the sample over by 180$\degree$ introduces misalignments between the phase images. These misalignments were corrected by using an automated Python script to identify matching features at the sample edges \citep{walt_scikit-image_2014,pena_hyperspyhyperspy_2019,paterson_fast_2020}. Linear distortions caused by the misalignment of the electron optical system of the TEM were corrected by using an affine transformation to provide an optimal match between the edges of each phase image pair, as shown in Fig.~\ref{fig_holo}c. In the present study, the images were found to be elongated by 2\% in a direction that forms a 75$\degree$ angle with the tilt axis by tilting a calibration sample by 180$\degree$ and tracking fiducial markers. By applying the affine transformation, phase separation artefacts were reduced to below the noise level, such that they were not detectable in line profiles. This approach ensures a consistent separation of $\varphi_{el}$ and $\varphi_{mag}$, as shown in the inset to Fig.~\ref{fig_holo}b. All necessary software packages are listed in Supplementary Material \suppl{[S3]}, and examples of intermediate steps of the hologram-to-phase calculation are shown in Supplementary Material \suppl{[S4]}.

\subsection{Alignment of phase images}\label{subsec23}
\textcolor{black}{As each $\varphi_{mag}$ image can have \mbox{non-rectangular} borders or contain pixels where the hologram-to-phase calculation resulted in errors, a confidence mask was used to identify the areas in each $\varphi_{mag}$ image that contained reliable measurements.}

\begin{figure*}[t]
        \centering\includegraphics[width=16cm]{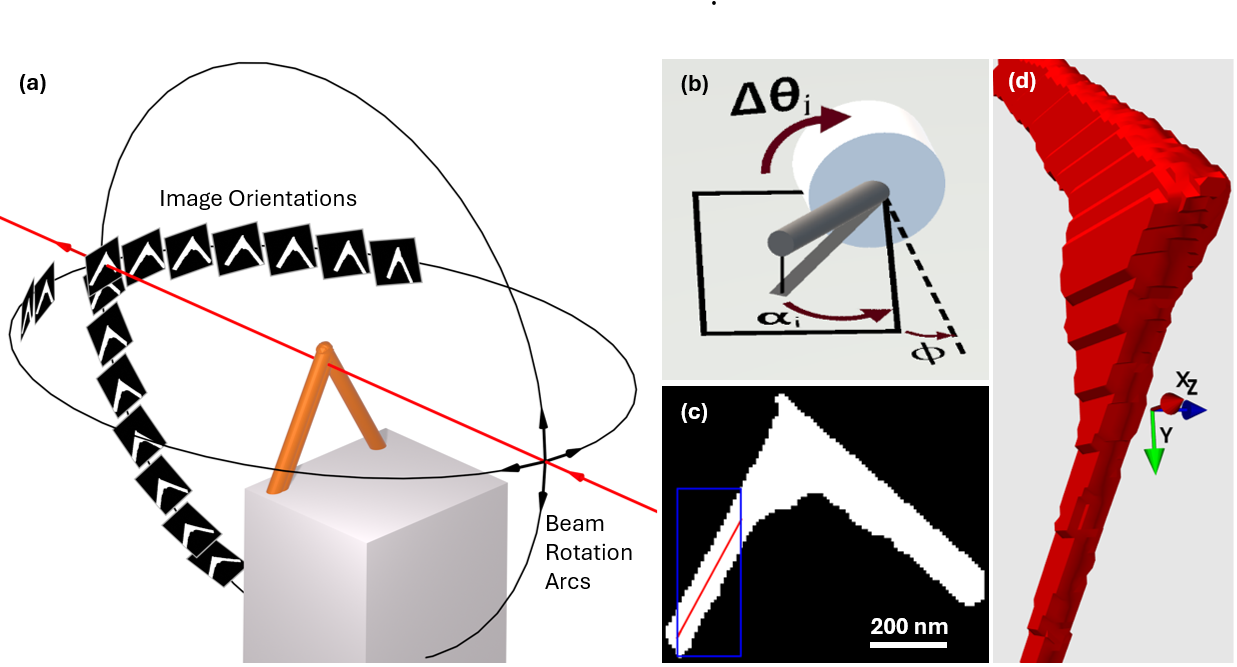}
        \caption{Alignment of tilt series of phase images and generation of a 3D geometrical model of the sample. (a)~Schematic diagram showing the angular position and sample outline of each phase image used in the reconstruction. The direction of the electron beam is shown in red. The arcs of rotation of the sample are shown in black. (b)~Rotations that define the projections for a conically-symmetrical element, showing the sample tilt $\Delta\theta_i$, the projection direction $\alpha_i$ and the rotation between the detector and the tilt axis $\phi$. (c)~The centre of an axially-symmetrical part of the sample can be identified in each projection and used to align the projections. (d)~3D tomographic reconstruction of the masks in the aligned projections performed to generate a model representing the sample geometry. A \mbox{10.2~nm} voxel size was used \textcolor{black}{because features smaller than \mbox{10~x~10~x~10~nm} cannot be resolved in the present study, in part due to tomographic misalignment. The use of a smaller voxel size would not improve the spatial resolution.}}
        \label{fig_tomo}
\end{figure*}

Each $\varphi_{el}$ image was first flattened to remove background ramps resulting from effects such as electron-beam-induced charging and the perturbed reference wave. A threshold was then used to define a mask that identified where the sample is located at each tilt angle. If elemental mapping was performed, as shown in Fig.~\ref{fig_sample}e, then the threshold could be fine-tuned to include only ferromagnetic material in the mask. These binary masks were used for aligning the phase images in each tilt series. 
Since the present phase images have a spatial resolution of no better than \mbox{6 nm} due to the chosen hologram fringe spacing, tomographic reconstruction was found to be highly sensitive to small image misalignments. Furthermore, the sample orientation was not known precisely during acquisition, as only relative tilt angles between phase images $\Delta\theta_i$ were measured. The projection positions shown in Fig.~\ref{fig_tomo}a were therefore refined by using an axial-symmetry-based implementation of the common lines method \citep{penczek_common-lines_1996}. As illustrated in Fig.~\ref{fig_tomo}b, at every sample tilt angle $\Delta\theta_i$ there is a unique projection orientation $\alpha_i$, and the two are related \emph{via} geometrical parameters: $\phi$ is the observed detector rotation relative to the sample tilt axis; $\theta_0$ is the starting tilt of the sample; $\alpha_0$ is the sample orientation at 0$\degree$ tilt. $\alpha_i$ can be measured by finding the sample symmetry axis from the electrostatic phase maps, as shown in Fig.~\ref{fig_tomo}c. The equation

\begin{equation}
\tan{\left(\alpha_i+\phi\right)}=\ \tan{\left(\alpha_0+\phi\right)}\cos(\theta_0+\Delta\theta_i)  		
\label{eq4}
\end{equation}
is then true for all projections and can be solved to find unknown constants ($\alpha_0$,$\ \phi$, $\theta_0$) defining the absolute sample orientation. For a single tilt series $i\ \in[1,\ldots,n]$ and, if the number of projections $n~>~3$, the system of equations is over-defined and can be solved numerically by using multivariate minimisation \citep{virtanen_scipy_2020}. Once they have been aligned, the electrostatic phase images can be back-projected to compute a scalar computed tomography (CT) reconstruction, as shown in Fig.~\ref{fig_tomo}d, which defines the geometry of the sample. In order to correct for missing wedge artefacts, the geometrical model (Fig.~\ref{fig_tomo}d) can be cropped to match the sample dimensions observed in the SEM images (Fig.~\ref{fig_sample}b). The refined model is used in the reconstruction of $\vect{M}$ to define where the magnetic material is located. As the alignment algorithm is imperfect, blurring of features is observed and the 3D spatial resolution of the CT reconstruction is poorer than the 6~nm 2D resolution of the phase images. As using a smaller voxel size did not improve the resolution of $\vect{M}$, the reconstruction uses a 10.2~nm voxel size.

\subsection{Three-dimensional reconstruction of \vect{M}}\label{subsec24}
In order to reconstruct the 3D distribution of \vect{M}, a forward model was used to simulate tilt series of magnetic phase images based on the geometrical 3D model of the sample (Fig.~\ref{fig_tomo}d) \citep{caron_chapter_3}. In order to determine the best-fitting values of \vect{M} in the model that would simultaneously satisfy all of the magnetic phase measurements, a cost function that could be minimised to determine the closest fit was defined according to the expression
\begin{equation}
\begin{gathered}
C=\sum_{i}\left(\varphi_{i,meas}-\varphi_{i,sim}(\vect{M})\right)^2 \\ +~\lambda_1\sum_{j}(\vect{\nabla} M_j \cdot \vect{\nabla} M_j)+\lambda_2 \text{var}(|\vect{M}|)~, 	
\label{eq5}
\end{gathered}
\end{equation}
where $C$ is the cost to minimise, $M_j$ are the magnetisation vector field components, $\varphi_i$ are the phase image pixels, $\lambda$ are the regulariser weights, $\vect{\nabla}$ is the gradient operator and "$\text{var}$" is the variance operator. As there are multiple equivalent solutions, regularising terms were added to favour solutions that exhibit a low gradient and a low variance in the magnitude of \vect{M}. The low gradient was chosen to favour ferromagnetic solutions, while the low variance was chosen to favour solutions comprising similar materials. The cost function was minimised by using a conjugate gradient method \citep{nocedal_numerical_2006} and initialised with $\vect{M}~=~0$. The regulariser weights ($\lambda$ in Eq.~\ref{eq5}) could be varied over 3 orders of magnitude and the reconstruction retained the same features, confirming that the solution was measurement-based. The reconstructed 3D distribution of $\vect{M}_{rec}$ was obtained for $\lambda_1~=~1$, which is the minimum value needed to lift the degeneracy between ferromagnetic and ferrimagnetic order, while $\lambda_2~=~0.1$ reduces surface artefacts without affecting the rest of the reconstruction.

\begin{figure*}[t]
        \centering\includegraphics[width=13cm]{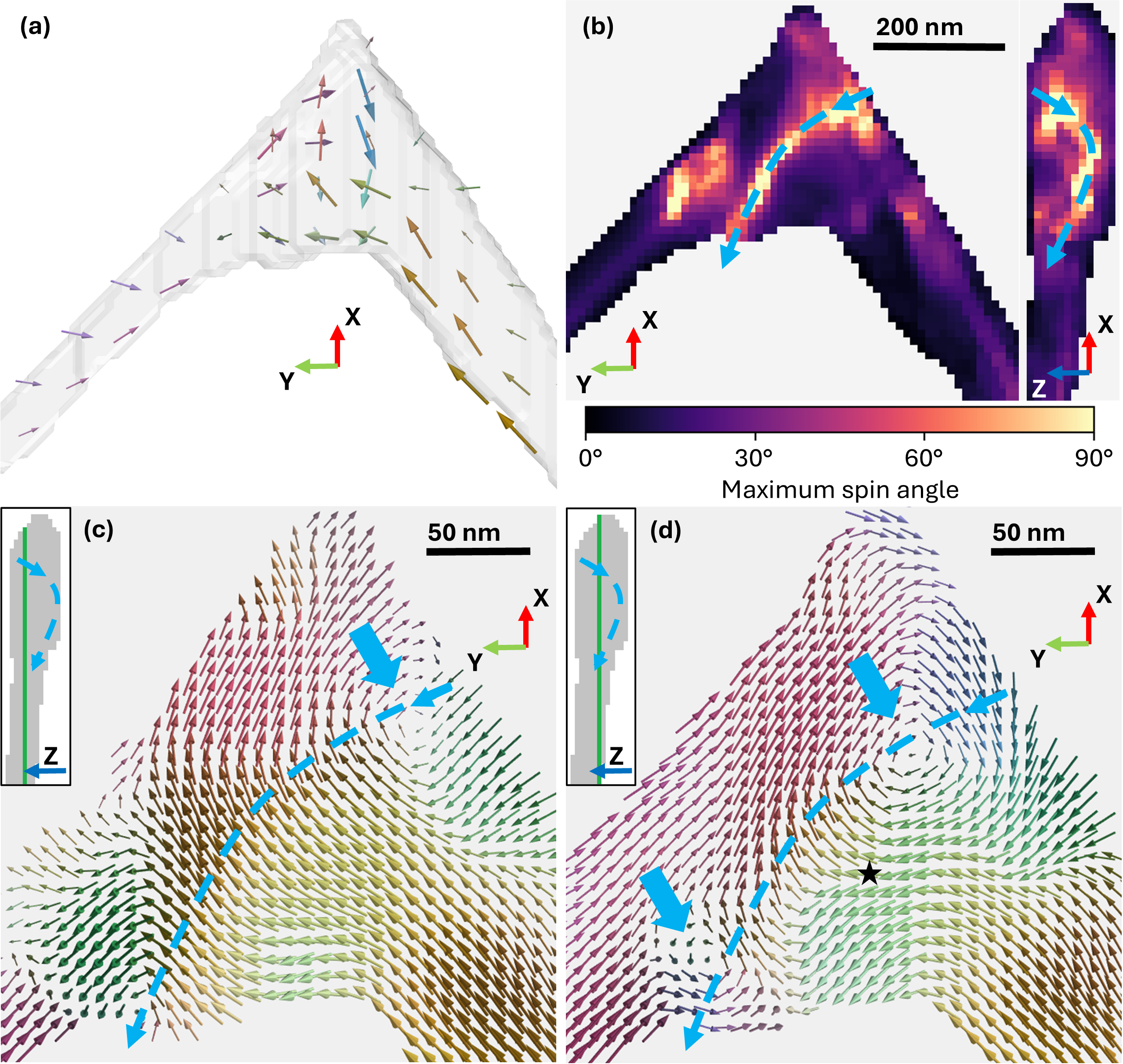}
        \caption{Reconstructed 3D magnetisation distribution $\vect{M}_{rec}$ in the cobalt nanostructure. The reconstruction is a 3D vector field, which maps the directions and magnitudes of the magnetic moments inside the sample. (a)~3D $\vect{M}_{rec}$ vector field. (b)~Maximum angle between neighbouring spins, projected along the \emph{z} and \emph{y} axes. The dashed lines follow the core of the vortex-type magnetic domain wall. (c,~d)~Single-plane slices of the $\vect{M}_{rec}$ vector field shown with full spatial resolution. The insets mark the locations of the slices. The vortex core is marked as in (b). Blue arrows mark where the slices intersect the vortex core, which intersects the top slice (c) once and the bottom slice (d) twice. The star in (d) denotes a voxel selected for diagnostics of the reconstruction.}
        \label{fig_reconstruction}
\end{figure*}

Fig. \ref{fig_reconstruction}a shows that the sample cotains multiple magnetic domains and that the magnitude of $\vect{M}_{rec}$ is greatest at the lower part of the intersection (denoted by the length of the arrows), in agreement with the radial cobalt distribution shown in Fig.~\ref{fig_sample}e. In order to view the positions of magnetic domain walls inside the structure, angles between pairs of neighbouring vectors were computed. The maximum angles are shown in Fig.~\ref{fig_reconstruction}b. An angle of $\sim 90\degree$ identifies the core of the vortex domain wall, while the dashed blue line shows the direction of the core, which corresponds to the $\text{curl}(\vect{M}_{rec})$ direction in sub-volumes around the vortex core. The magnetic vortex state occupies the full $\sim 200$~nm~x~200~nm~x~100~nm volume of the nanowire intersection. The U-shaped vortex core has a length of 350~nm. Fig.~\ref{fig_reconstruction}c-d shows $z$~slices extracted from $\vect{M}_{rec}$ at the full resolution of the reconstruction, containing intersections with the vortex magnetic domain wall. 

\subsection{Diagnostics of the reconstruction}\label{subsec25}
\begin{figure}[h]
        \centering\includegraphics[width=6cm]{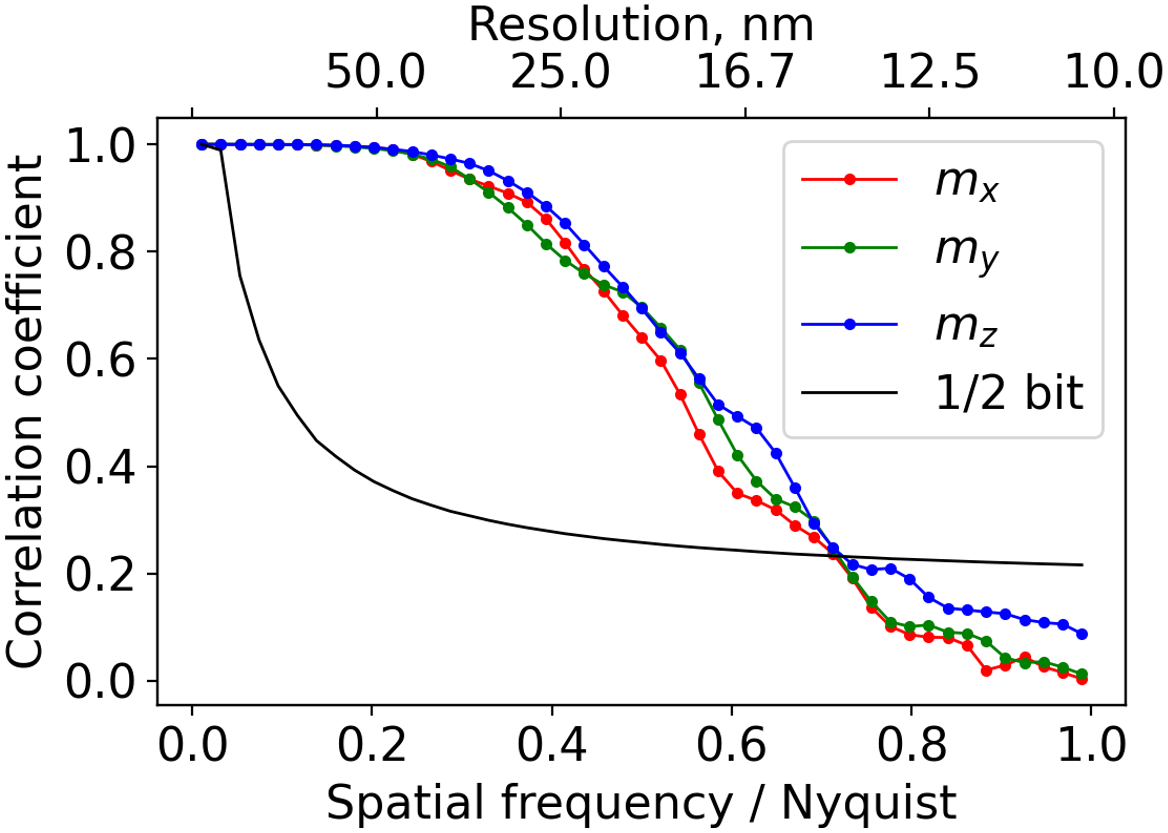}
        \caption{Fourier shell correlation evaluation, showing that reconstructed features larger than \mbox{14.8~nm} can be interpreted directly.}
        \label{fig_fsc}
\end{figure}

The inflluece of noise on the resolution of the reconstruction was assessed by using a Fourier shell correlation. This approach involved halving the 3D $\vect{M}_{rec}$ dataset through random sampling and interpolating the missing values. Fig.~\ref{fig_fsc} displays an estimate of the signal-to-noise ratio (SNR) per spatial frequency band in the 3D reconstruction volume. The correlation is plotted relative to the $\frac{1}{2}$ bit SNR threshold commonly used in X-ray tomography \citep{van_heel_fourier_2005}. It reveals that any feature larger than 14.8~nm has sufficient SNR to be interpreted directly. The resolution limit of $\sim 1.5$~pixels due to noise is regarded as conservative, since noise is suppressed in regularised reconstructions. Hence, systematic uncertainties are also considered. If the angle between neighbouring atomic spins is larger than 30$\degree$, then the micromagnetic energy expressions are no longer valid \citep{donahue_exchange_1997}. The maximum spin angle in Fig.~\ref{fig_reconstruction}b corresponds to the positions of domain walls and suggests that the calculations are valid in most of the sample, although voxels at the core of the vortex are not resolved.

\begin{figure*}[t]
        \centering\includegraphics[width=12.5cm]{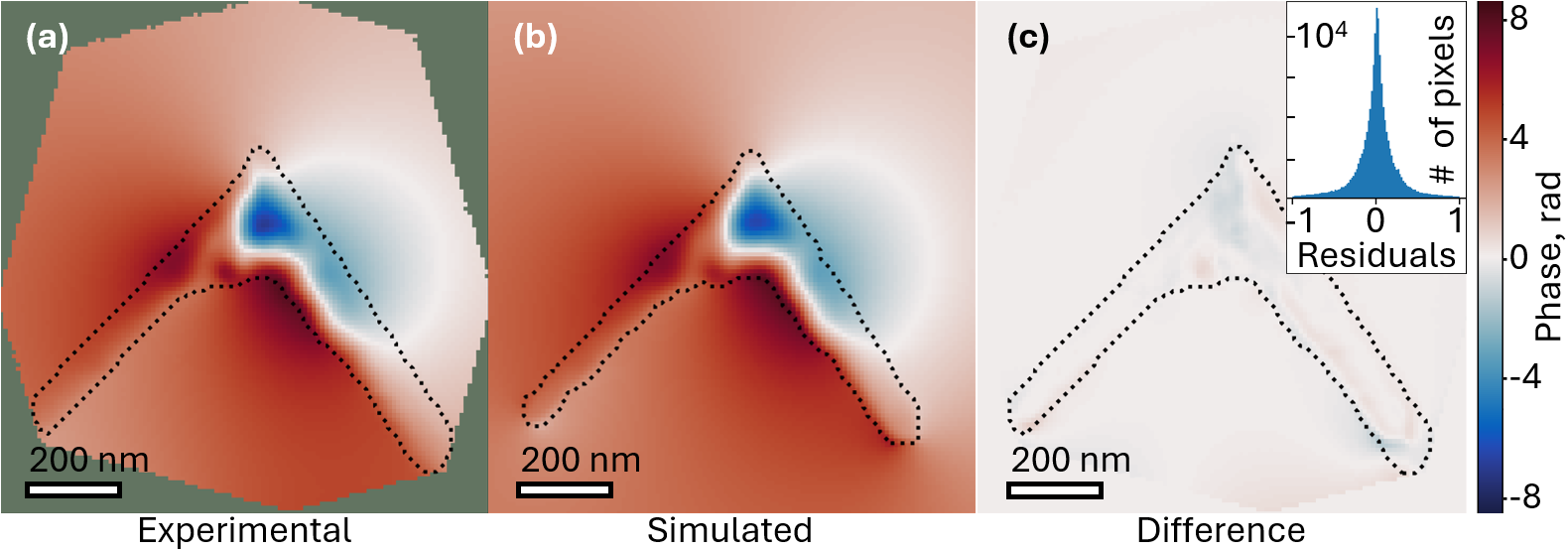}
        \caption{Magnetic phase shift reconstruction for a sample tilt angle of -3.7$\degree$. (a)~Experimental magnetic phase image. (b)~Magnetic phase image calculated from the reconstructed magnetisation distribution. (c)~Difference between (a) and (b). The inset shows the distribution of residual differences for magnetic phase images at all tilt angles. The residuals are used to estimate the $\vect{M}_{rec}$ reconstruction uncertainty.}
        \label{fig_phase_diff}
\end{figure*}

\begin{figure}[ht]
        \centering\includegraphics[width=6.5cm]{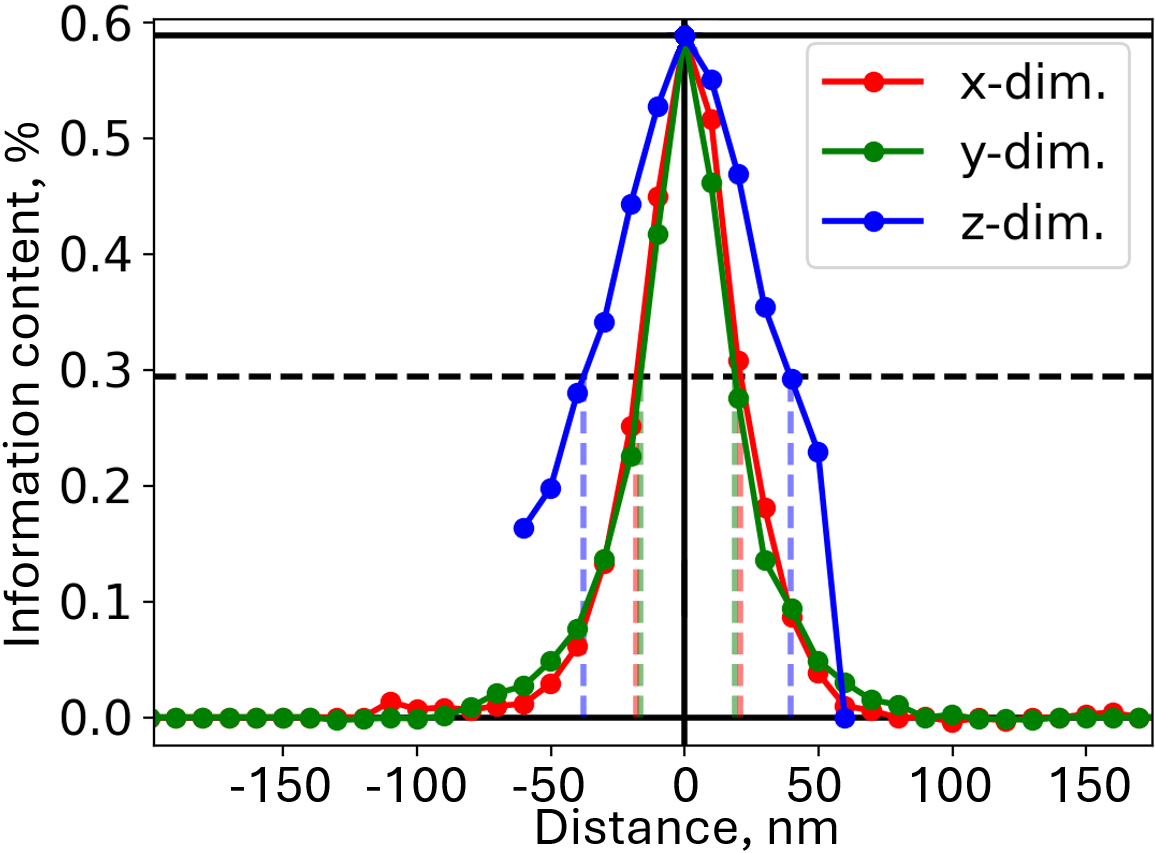}
        \caption{3D point spread function, showing the spatial distribution of information content for the voxel denoted by a star in Fig.~\ref{fig_reconstruction}d. For the present dataset and reconstruction procedure, the information contained at this point in space is spread over a volume with an average FWHM of 43~nm, in part due to computational limitations.}
        \label{fig_fwhm}
\end{figure}

Other uncertainties can be identified by using optimal estimation diagnostics \citep{ungermann_towards_2010}, by assuming that all \vect{M} vectors and phase image pixels have probability density functions that can be described by a normal distribution. The reconstructed magnetisation $\vect{M}_{rec}$ can be expressed as the true distribution of \vect{M} transformed by an averaging kernel matrix $A$, which is further affected by measurement errors  $\epsilon$, according to the error gain matrix $G$:
\begin{equation}
    \vect{M}_{rec}=A \vect{M}+G\epsilon 	~.
    \label{eq6}
\end{equation}
$A$ and $G$ were determined for the $\vect{M}_{rec}$ vector denoted by a star in Fig.~\ref{fig_reconstruction}d, which has $\mu_0 M_s~=~0.78$~T and is well resolved with a maximum spin angle of 13$\degree$. $G$ is used for linear error propagation because it expresses the error on a single $\vect{M}_{rec}$ vector as a function of error on every pixel in experimental phase measurements. The phase measurement noise of 0.016~rad per pixel was determined by calculating the standard deviation of the vacuum measurements and is mapped by $G$ to a $\mu_0 M_s$ error of 0.001~T. The experimental and simulated phase images used in the cost function in Eq.~\ref{eq5} cannot match perfectly due to experimental misalignments, resulting in a residual distribution \mbox{Fig.~\ref{fig_phase_diff}} corresponding to a root mean square error of 0.38~rad. Forcing the reconstruction to match any one image perfectly would, on average, introduce a perturbation that maps to a $\mu_0 M_s$ error of 0.009~T. $A$ is a measure of spatial averaging during the reconstruction and shows that, in order to calculate a single $\vect{M}_{rec}$ vector, information is taken from a volume around the central position, as illustrated in \mbox{Fig.~\ref{fig_fwhm}}. For this voxel, the point spread function has an average full-width-at-half-maximum (FWHM) of 43~nm. The evaluation of A is equivalent to changing the value of one voxel of $\vect{M}_{rec}$ and repeating the reconstruction to quantify systematic errors. As error propagation and perturbation mapping for other voxels yields similar results, the $\mu_0 M_s$ measurement precision is determined to be 0.01~T per pixel and the spatial resolution to be better than 50~nm. 

For the present dataset, it would have been possible to perform the reconstruction with a voxel size of 5.1~nm instead of 10.2~nm. However, smaller voxels would have been affected more strongly by noise and phase image misalignments, resulting in reduced precision per pixel. During reconstruction, gradient regularisation is needed to obtain ferromagnetic solutions. However, it also suppresses noise, applies smoothing to sharp features, and can be adjusted to vary the trade-off between precision and spatial resolution. If the precision for smaller voxels were to be increased to 0.01~T by adjusting the regulariser weights, it would result in a similar spatial resolution, which is largely independent of voxel size for voxels smaller than 10~nm. This behaviour is attributed to remaining uncorrected phase image misalignments, which result in the smoothing of features in the reconstruction. By using a 10.2~nm voxel size, the MBIR algorithm computes a reconstruction in 3 hours using one core of an Intel i7 central processing unit (CPU). The algorithm optimisations are discussed in the Supplementary Material \suppl{[S5]}. Achieving a higher resolution would require more advanced image drift and distortion correction, optimised reconstruction software that computes faster, and the cumulative acquisition of larger datasets.

\section{Results}\label{section3}
\subsection{Uniform region reconstruction}\label{subsec31}

\begin{figure}[ht]
        \centering\includegraphics[width=6.5cm]{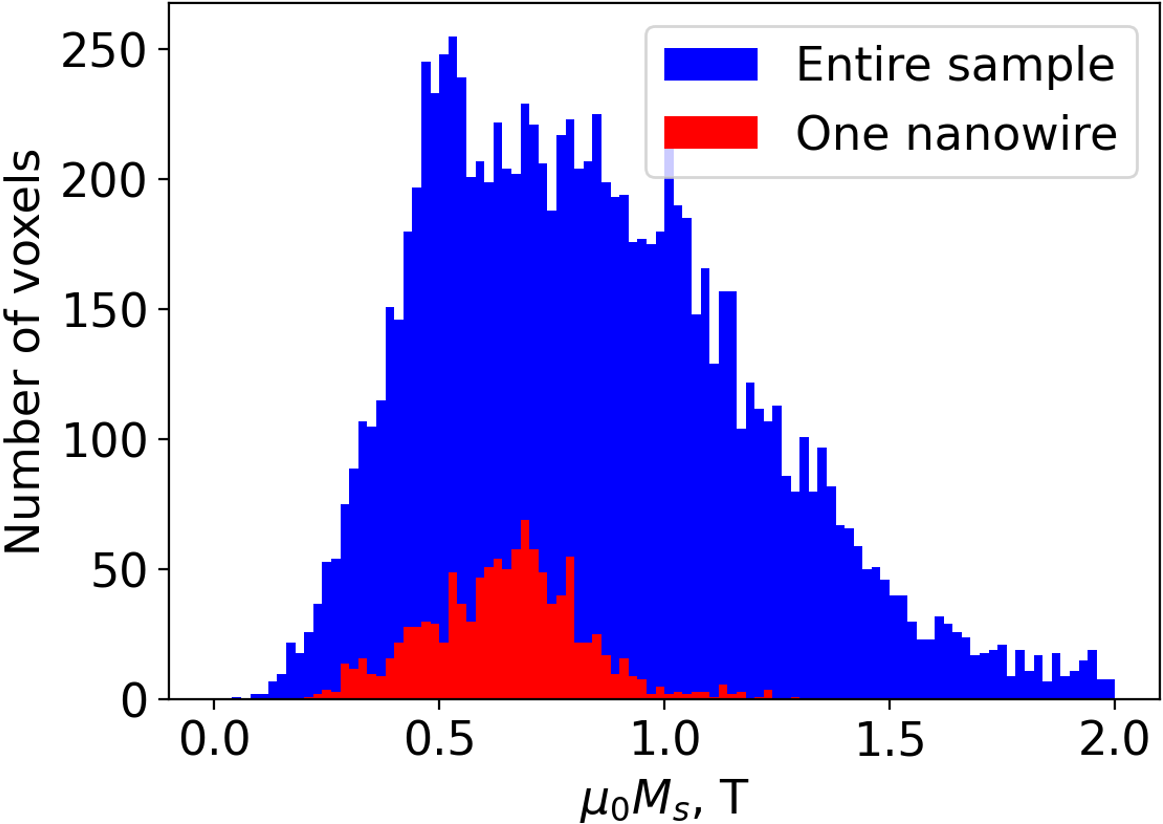}
        \caption{Distribution of magnitudes of the reconstructed magnetisation distribution $\vect{M}_{rec}$, shown for all of the voxels in the sample in blue and for the voxels in the nanowire on the left side of Fig.~\ref{fig_reconstruction}a in red. The bin size is 0.02~T. The distributions result from material composition variations and reconstruction artefacts.}
        \label{fig_hist}
\end{figure}

As described in the preceding Section, EH-VFT and MBIR were used to record and reconstruct the 3D distribution of $\vect{M}_{rec}$ in a sample comprising two intersecting nanowires. Fig.~\ref{fig_sample}e shows that the nanowires contain a mixture of cobalt and organic compounds, which transitions from a 60\% cobalt content in the core to a pure carbon shell at the surface. Since the cobalt is dispersed within the carbon matrix in nanocrystalline form \citep{pablo-navarro_purified_2018} and bulk cobalt has \mbox{$\mu_0 M_s~=~1.75$~T} \citep{pablo-navarro_purified_2018}, the nanowires should exhibit $\mu_0 M_s~<~1.1 $~T, as observed for the nanowire distribution in \mbox{Fig.~\ref{fig_hist}.} The few outliers with \mbox{$\mu_0 M_s~>~1.1$~T} are likely to be artefacts at the interface between the magnetic material and the vacuum region, where magnetic voxels that are designated as non-magnetic force the reconstruction to increase the value of $M_s$ locally. In other parts of the sample, surface voxels contain non-magnetic material where $M_s$ is close to zero. Surface reconstruction artefacts are present in part because missing wedge correction for the geometrical model is not perfect, as discussed in the Supplementary Material \suppl{[S1]}. The threshold for creating the 3D geometrical model shown in Fig.~\ref{fig_tomo}d was refined after the first reconstruction, so that similar amounts of over- and underestimating surface artefacts were present. \textcolor{black}{A discussion of the threshold for the geometrical model and a line trace showing surface artefacts are included in Supplementary Material \suppl{[S6]}.} Such surface artefacts are also identifiable for the full sample as voxels with $\mu_0 M_s~>~1.75$~T, as shown in the histogram in \mbox{Fig.~\ref{fig_hist}.} Such errors could be reduced in the future by using smaller voxels and improved tomographic alignment and distortion removal algorithms. Both nanowires have single magnetic domain configurations at a distance of 200~nm from the intersection, with $\mu_0 M_s~=~0.7 \pm 0.2$~T in the left nanowire and $\mu_0 M_s~=~0.5 \pm 0.2$~T in the right nanowire. $M_s$ has previously been measured for varying cobalt contents of FEBID nanowires \citep{teresa_review_2016,pablo-navarro_purified_2018}, but the measurements were corrected to exclude the non-magnetic shell. The latter correction can be replicated by assuming that in Fig.~\ref{fig_sample}e the compositional line profile corresponds to a circular-cross-section nanowire with a 100-nm-diameter magnetic core, resulting in an estimated average cobalt content of 61\% with 4\% standard deviation, which would correspond to $\mu_0 M_s~=~0.5 \pm 0.1$~T. The reconstructed value of $\vect{M}_{rec}$ is consistent with the predicted values, but shows a higher standard deviation due to surface artefacts.

\subsection{Multi-domain reconstruction}\label{subsec32}
The primary advantage of reconstructing $\vect{M}_{rec}$ is to allow the visualisation of local variations in magnetic domains and the twisting of magnetic domain walls. Fig.~\ref{fig_reconstruction}a and Fig.~\ref{fig_sample}c are projected along the $z$~axis. The close-to-anti-parallel $\vect{M}_{rec}$ vectors in Fig.~\ref{fig_reconstruction}a explain the phase shift variations at the nanowire intersection in Fig.~\ref{fig_sample}c. The $\text{curl}(\vect{M}_{rec})$ direction in Fig.~\ref{fig_reconstruction}b shows that in Fig.~\ref{fig_sample}c only part of the vortex has its curl oriented close to the $z$~axis, while the other part is more in-plane and requires imaging about a second tilt axis for a sufficient signal to be detected. The magnetic vortex core is curved to favour its positioning at the protrusion at the bottom of the intersection, which is considered to be a domain wall pinning site. The limited spatial resolution is evident in Fig.~\ref{fig_reconstruction}c, where the vortex core has a lower value of $M_s$ than the surrounding material due to spatial averaging of vector components of opposite sign. This averaging is a result of the regularisation in Eq.~\ref{eq5}, in which the preferential selection of solutions with low gradients introduces smoothing. Although this smoothing is necessary to constrain solutions for $\vect{M}_{rec}$  to retain ferromagnetic ordering, the averaging of anti-parallel vectors offsets $M_s$ values at the vortex core. In the present experiment, the $M_s$ offset due to spatial averaging is negligible when compared to variations in material composition. The elemental map shown in Fig.~\ref{fig_sample}d indicates that the cobalt content at the intersection is elevated to $70 \pm 15$\%, with a predicted $\mu_0 M_s~=~0.8 \pm 0.5$~T \citep{teresa_review_2016, pablo-navarro_purified_2018, donnelly_complex_2022}, in close agreement with the value of $\mu_0 M_s~=~0.9 \pm 0.3$~T measured from the reconstruction. \textcolor{black}{The correlation between $M_s$ and the atomic Co content is consistent with previous work, as shown in Supplementary Material \suppl{[S6]}.} The magnetic domain wall occupies the full \mbox{box$\sim 200$~nm} width of the intersection, which is comparable to the $\sim 100$~nm magnetic domain wall width of transverse head-to-head domain walls that previously observed using magnetic X-ray laminography in 80\% purity FEBID cobalt nanowire double helixes \citep{donnelly_complex_2022}. In summary, MBIR has provided accurate measurements of $M_s$ for a U-shaped vortex magnetic domain wall of width 200~nm,  but is less sensitive to sub-50 nm $M_s$ variations in the present study due to limitations in spatial resolution and sensitivity that can be overcome in future experiments.

\section{Discussion}\label{section4}
\subsection{Algorithm improvements}\label{subsec41}

The 3D spatial resolution of the reconstruction could be improved significantly by reducing the voxel size, as this would reduce the extent of spatial averaging. Since imposing ferromagnetic ordering requires a comparison of neighbouring voxels, the resolution is limited to approximately three times the voxel size. Therefore, a  3D resolution of 3~nm would be possible if a 1~nm voxel size were used. Such a resolution improvement would increase the computation time by a factor of one thousand. However, the MBIR algorithm can be parallelised to benefit from computing on dedicated servers and conjugate gradient minimisation could be replaced by a linear least squares solver to reduce the number of minimisation steps if the variance regularisation in \mbox{Eq.~\ref{eq5}} was removed \citep{virtanen_scipy_2020}. The current symmetry-based tomographic alignment shows up to 10~nm image alignment errors when applied to samples comprising more than a single cylindrical nanowire. More general algorithms would be suitable for other samples \citep{penczek_common-lines_1996, houben_refinement_2011}. Off-axis electron holography can provide measurements of phase shift with sub-nm spatial resolution by reducing the interference fringe spacing on a dedicated electron microscope \citep{dunin-borkowski_electron_2019}, ultimately with atomic spatial resolution in magnetic-field-free conditions \citep{tanigaki_electron_2024}. 

\subsection{Outlook}\label{subsec42}
The present study provides a quantitative reconstruction of the 3D distribution of \vect{M} for a cobalt nanostructure using EH-VFT. Similar datasets have previously been used to reconstruct 3D distributions of \vect{B} \citep{prabhat_3d_2017, wolf_holographic_2019, lewis_wrap_2023}. 3D reconstructions of \vect{M} were tested for a range of nanowires of varying purity. A nanowire intersection was chosen for this paper, as it combines the complexity arising from the non-uniformity of the material with the possibility of projection-based null spaces from a 3D vortex magnetic domain wall. The reconstruction is shown to be accurate based on 16 projections with a 10$\degree$ angular spacing over two tilt axes with sub-50-nm spatial resolution. For other nanostructures, the reconstruction of \vect{M} is likely to be possible if the following conditions are met: 
\begin{itemize}
    \item The sample geometry must be defined. Errors arising from an inaccurate geometrical model are discussed in the Supplementary Material \suppl{[S1]}.
    \item The phase shift arising from all components of \vect{M} must be measured. In general, two complete tilt series are necessary. In the case of uniform nanowires and vortex states, lower tilt ranges are viable, but the reconstruction error is then increased, as discussed in the Supplementary Material \suppl{[S1]}.
    \item There should ideally not be external magnetic fields, or conduction or displacement currents, in the sample, in order to ensure that the measured phase shift is solely due to \vect{M}.
    \item The sample must be ferromagnetic. This is a requirement for the MBIR algorithm.
\end{itemize}
If null spaces are present, then they only affect the corresponding sub-volume, as discussed in the Supplementary Material \suppl{[S2]}. The ferromagnetic nanostructure shown in this paper satisfies the above conditions because the sample dimensions were measured during fabrication and were used to correct for  missing wedge artefacts in the geometrical model, while the configurations of \vect{M} that are present in the sample do not create null spaces and can, in principle, be reconstructed with an error of less than 1\% when imaged over $\pm 60 \degree$ tilt arcs. I addition, the sample is isolated from other material that may accumulate charge or be magnetised.

Previous works have shown 2D reconstructions of \vect{M} \citep{kovacs_mapping_2017, song_quantification_2018}, while an alternative MBIR implementation \citep{mohan_model-based_2018} has provided only a qualitative 3D reconstruction. Micromagnetic simulations \citep{wolf_unveiling_2022, lyu_three-dimensional_2024} are a common approach to 3D quantification, but require a knowledge of the magnetic properties of the material, whereas MBIR can, in special cases, reconstruct \vect{M} without such prior knowledge. Quantitative measurements have been reported using both X-ray and neutron tomography \citep{donnelly_three-dimensional_2017, hilger_tensorial_2018, donnelly_time-resolved_2020, donnelly_complex_2022, di_pietro_martinez_three-dimensional_2023}, with a spatial resolution of 50~nm for X-ray laminography. \textcolor{black}{Our MBIR results are consistent with \mbox{X-ray} tomography results and micromagnetic simulations of similar FEBID nanostructures \citep{donnelly_complex_2022}.} \textcolor{black}{When experimental and computational limitations are resolved to improve} spatial resolution, MBIR would be uniquely suited to characterise 3D magnetic nanostructures, such as nanowire junctions, interfaces in nanoparticles, Bloch-type skyrmions on curved surfaces, and hopfions. In total, this method reconstructs projections of \vect{B}, projections of electrostatic potential, 3D distributions of \vect{M}, and 3D geometry of nanostructures.

\section{Conclusions}\label{section5}
Model-based iterative reconstruction has been applied and further developed to measure the 3D distribution of magnetisation \vect{M} in a FEBID ferromagnetic nanostructure from multiple projections of the magnetic induction field recorded using off-axis electron holography. The effect of null space limitations on the uniqueness and accuracy of the reconstruction has been assessed. \textcolor{black}{The results are in agreement with X-ray tomography results obtained from similar structures.} Optimal estimation diagnostics show that the reconstruction has a precision of 0.01~T at each voxel. The present spatial resolution of several tens of nanometres is restricted primarily by computational limitations and can be improved greatly in future experiments (towards the atomic scale) by using a smaller voxel size, more specialised electron microscopes, and improved image processing.

\section{Availability of data and materials}
\textcolor{black}{The data underlying this article are available in Enlighten at \href{https://doi.org/10.5525/gla.researchdata.1953}{https://doi.org/10.5525/gla.researchdata.1953}.}

\section{Acknowledgements}
The authors are grateful to Jan Caron for the theoretical and computational development of \vect{M} reconstruction using MBIR and to Patrick Diehle for experimental contributions to the development of EH-VFT.

\section{Funding}
The authors are grateful for funding from the Engineering and Physical Science Research Council (Grant No. EP/W524359/1 and EP/X025632/1), the European Research Council under the European Union's Horizon 2020 Research and Innovation Programme (Grant No. 856538, project ``3D MAGiC''), the Deutsche Forschungsgemeinschaft (Project-ID 405553726 TRR270) and a Helmholtz fellowship.

\section{Conflict of interest}
The authors declare none.

\section{Author contributions}
A\v{S}: Software, Formal analysis, Writing -- original draft. AK: Supervision, Methodology, Writing -- review \& editing. SMV: Writing -- review \& editing. REDB: Writing -- review \& editing. KF: Writing -- review \& editing. TPA: Supervision, Methodology, Writing -- review \& editing.

\bibliographystyle{unsrtnat}
\bibliography{mam-authoring-template}
\end{document}